\documentclass[]{aa}
\usepackage{graphicx}
\usepackage{txfonts}
\usepackage{natbib}
\bibpunct{(}{)}{;}{a}{}{,}
\usepackage{color}

\begin{document}

\title{A Sino-German $\lambda$6~cm polarisation survey of the Galactic plane} 
\subtitle{VIII. Small-diameter sources}

\author{W. Reich\inst{1}
        \and X. H. Sun\inst{1,2,3} 
        \and P. Reich\inst{1}
        \and X. Y. Gao\inst{2}
        \and L. Xiao\inst{2}
        \and J. L. Han\inst{2}
        }

\institute{Max-Planck-Institut f\"{u}r Radioastronomie, 
            Auf dem H\"ugel 69, 53121 Bonn, Germany\\
            \email{wreich@mpifr-bonn.mpg.de}
          \and National Astronomical Observatories, CAS, Jia-20 Datun Road, 
           Chaoyang District, Beijing 100012, China
         \and Sydney Institute for Astronomy, School of Physics, The University of 
           Sydney, NSW 2006, Australia
             }

\date{Received / Accepted}

\abstract
{} 
{Information of small-diameter sources is extracted from the Sino-German $\lambda$6~cm polarisation survey 
of the Galactic plane carried out with the Urumqi 25-m telescope.
}
{We performed two-dimensional elliptical Gaussian fits to the $\lambda$6~cm 
maps to obtain a list of sources with total-intensity and polarised flux densities. 
}
{The source list contains 3832 sources with a fitted diameter smaller than $16\arcmin$ and
a peak flux density exceeding 30~mJy, so about 5$\times$ the rms noise, of the total-intensity data.
The cumulative source count indicates completeness for flux densities exceeding about 60~mJy. 
We identify 125 linearly polarised sources at $\lambda$6~cm with a peak polarisation flux 
density greater than 10~mJy, so about 3$\times$ the rms noise, of the polarised-intensity data. 
} 
{Despite lacking compact steep spectrum sources, the $\lambda$6~cm catalogue lists about 20\% more sources 
than the Effelsberg $\lambda$21~cm source catalogue at the same angular resolution and for the same area. 
Most of the faint $\lambda$6~cm sources must have a flat spectrum and are either $\ion{H}{II}$ regions 
or extragalactic. When compared with the Green Bank $\lambda$6~cm (GB6) catalogue, we obtain higher flux densities 
for a number of extended sources with complex structures. Polarised $\lambda$6~cm sources density 
are uniformly distributed in Galactic latitude. Their number density decreases towards the inner Galaxy. 
More than 80\% of the polarised sources are most likely extragalactic. With a few exceptions, the sources 
have a higher percentage polarisation at $\lambda$6~cm than at $\lambda$21~cm. Depolarisation seems to 
occur mostly within the sources with a minor contribution from the Galactic foreground emission.
}

\keywords{radio sources - polarisation 
}

\titlerunning{A Sino-German $\lambda$6~cm polarisation survey of the Galactic plane: VIII. Small-diameter sources}
\maketitle

\section{Introduction}

Diffuse continuum emission is highly concentrated in a narrow band along
the Galactic plane, where the number density of discrete Galactic sources, such as
supernova remnants (SNRs) and $\ion{H}{II}$ regions, is the highest. Recently,
the maps from the Sino-German $\lambda$6~cm polarisation survey of the Galactic 
plane have been published in a series of papers \citep{shr+07,grh+10,srh+11,xhr+11}. 
The survey also served as one basis for a systematic study of known SNRs \citep{srr+11,gsh+11a}
and for a search for new ones \citep{ghr+11b}. \citet{hrs+13} summarise the
$\lambda$6~cm survey project and the results obtained so far.
This paper adds a catalogue of discrete small-diameter sources in the surveyed area, including
polarised sources.

At $\lambda$6~cm, optically thin diffuse thermal emission dominates in the Galactic plane,
while diffuse steep-spectrum synchrotron emission dominates at longer wavelengths. 
Thus, faint $\ion{H}{II}$ regions are expected to be better separable and therefore identified 
from the diffuse emission at $\lambda$6~cm  
than in the single-dish surveys carried out with the Effelsberg 100-m telescope at 
$\lambda$21~cm \citep{rrf+90,rrf+97} and $\lambda$11~cm \citep{rfs+84,rfr+90,frr+90a}. 
On the other hand, the extraction of faint steep-spectrum 
extragalactic sources from the diffuse Galactic emission becomes more
difficult at $\lambda$6~cm. In any case, $\lambda$6~cm flux densities are a valuable addition
to existing longer wavelength data, and a more precise spectral index determination 
is provided when the wavelength difference increases.

Synthesis telescope source surveys, such as the most sensitive NRAO VLA Sky Survey (NVSS) 
\citep{ccg+98} at $\lambda$21~cm,
list many more compact sources along the Galactic plane compared to published single-dish surveys. 
These observations have much higher angular resolution and sensitivity. They filter out
extended Galactic emission and also may underestimate the integrated flux density of extended sources.

Previously, an important $\lambda$6~cm source survey of the northern sky, the GB6, had been 
carried out with the former Green Bank 300-ft telescope with an angular resolution 
of $3\farcm6 \times 3\farcm4$ by \citet{cbs+94}. The corresponding source catalogue \citep{gsd+96} 
includes compact sources up to $10\farcm5$. The peak flux density   
increases with declination with a lower limit of 18~mJy \citep{gsd+96}. In the reduction 
process of the GB6 survey, 
spline functions were used to subtract all kind of extended emission from the raw data, 
which includes Galactic emission. Therefore the maps in the Galactic plane can not be 
compared with those from the present $\lambda$6~cm survey. The GB6 survey processing should not 
affect the flux density determination of compact sources, what can be proved by comparison 
with the present catalogue. 
 
Lists of polarised sources in the Galactic plane are rare and do not exist at all at 
$\lambda$6~cm. For $\lambda$21~cm, however, several data sets exist: The 
NVSS \citep{ccg+98} includes polarisation information. \citet{btj+03} provides polarisation 
data from the Canadian Galactic Plane Survey (CGPS) and \citet{vbs+11} presents a list of polarised sources 
along the Galactic plane measured with the VLA, which could be compared with the 
$\lambda$6~cm polarisation data.
The Effelsberg $\lambda$21~cm and $\lambda$11~cm survey source lists (available from 
the CDS/Strasbourg\footnote{http://vizier.u-strasbg.fr}) have guided us in the layout of 
the $\lambda$6~cm source list, where all relevant parameters of individual sources 
are provided. In Sect.~2, we summarise the $\lambda$6~cm survey project and describe
the source fitting procedure in Sect.~3. In Sect.~4, we present the list of sources and 
describe their individual parameters. In Sect.~5, we briefly discuss the $\lambda$6~cm source 
catalogue with respect to source statistics and other available catalogues and make 
concluding remarks in Sect.~6.

\section{The $\lambda$6~cm survey}

The Sino-German $\lambda$6~cm polarisation survey of the Galactic plane was conducted with the 
Urumqi 25-m telescope of Xinjiang (formerly Urumqi)
Astronomical Observatory, Chinese Academy of Sciences, between 2004 and 2009.
The surveyed area covers $10\degr \le \it{l} \le$ 230$\degr$ and 
$-5\degr \le \it{b} \le$ +5$\degr$. The survey has an angular resolution of $9\farcm5$. 
The system temperature  towards the zenith was about 22~K. The central 
frequency was set to either 4.8~GHz or 4.963~GHz with 
corresponding bandwidths of 600~MHz and 295~MHz. The system gain is 
$\rm T_{B} [K]/S[Jy]$ = 0.164. Detailed information about the 
receiving system, the survey set-up, and its reduction scheme has already been presented by 
\citet{shr+07}. The survey maps were published in three sections by 
\citet{grh+10,srh+11, xhr+11}
and are also available on the web\footnote{http://zmtt.bao.ac.cn/6cm/}.

The Galactic plane was fully sampled at $3\arcmin$ and mapped by raster scans in 
the longitude and latitude directions. 
The primary calibrator was 3C~286 with an assumed flux density of 7.5~Jy and a
polarisation percentage of 11.3\%. The polarisation angles measured for 3C~286 
were found as 32$\degr \pm 1\degr$ and were not corrected to the nominal value 
of 33$\degr$. Then 3C~48 and 3C~138 were used as secondary calibrators, and 3C~295 and 3C~147 as 
unpolarised calibrators. \citet{shr+07} found a scaling accuracy of better than 4\%
for total intensities and 5\% for polarised intensities. 

The raw data from the receiving system contain maps of $I$, $U$, and $Q$ stored 
in \textsc{NOD2}-format \citep{has74}. Data processing follows the standard 
procedures developed for continuum observations with the Effelsberg 100-m 
telescope as detailed by \citet{shr+07} and \citet{grh+10}. The positional
accuracy of compact sources in the survey maps was found to be in general
better than $1\arcmin$ when compared with the high-resolution interferometric 
NVSS survey \citep{ccg+98}. Maps with larger position offsets were corrected with respect
to the NVSS source positions. The final survey maps have 
a typical measured rms noise including confusion of about 1~mK~$T_{\rm B}$ or 6.1~mJy/beam area 
for total intensity $I$, 0.5~mK~$T_{\rm B}$ or 3.05~mJy/beam for Stokes 
$U$ and $Q$, and polarised intensity $PI$.  

\section{Source fitting procedure}

\subsection{Total intensity fit}

We used the same Gaussian fitting routine applied to extract compact sources from the Effelsberg 
$\lambda$21~cm \citep{rrf+90,rrf+97} and $\lambda$11~cm survey maps 
\citep{rfs+84,frr+90b} to produce a list of compact sources from
the $\lambda$6~cm survey maps. This is the standard NOD2-based fitting routine for 
continuum and polarisation observations with the Effelsberg 
100-m telescope, which has a Gaussian beam shape up to mm-wavelengths.     

The fitting routine can be steered in various ways. The standard procedure is to run an 
automatic fit on a map, where a small area around each
source is extracted and corrected for baseline gradients before a fit is applied, which is either
a circular or an elliptical Gaussian. The highest peaks in a map are fitted in a first run, 
and subsequently the peak amplitude limit is decreased. 
It turns out, that for most $\lambda$6~cm sources, the automatic procedure does not give the best result 
as seen by the residual emission after source subtraction, so that most sources were fitted 
individually from the maps by defining the area for the fit where confusing surrounding emission
is excluded as well as possible.    
 
The inner Galactic plane has steep intensity gradients in Galactic latitude, which makes it difficult to
extract faint sources. 
In analogy to the treatment of the Effelsberg survey maps, we applied the ``unsharp-masking'' 
filtering method by \citet{sr79} by using a $1\degr$ wide filtering beam to remove most of
the diffuse emission before applying the source fitting routine, which improves
the number of separable sources from unrelated emission and also improves the fit result. 
The peak flux amplitude limit was taken as 5$\times$ the rms noise, e.g. 5~mK~$T_{\rm B}$ 
or 30~mJy/beam area. In addition we rejected all source fits with the minor axis 
below $6\arcmin$, which indicates either an RFI-spike, another small-scale distortion,
or surroundings that are  too complex.
Also, fit results of the major axis exceeding $16\arcmin$ were rejected, which is the
same limit as used earlier for the Effelsberg $\lambda$21~cm source list with about the same beam size. 
For sources that are significantly larger than the angular resolution, the application of a single 
Gaussian fit becomes questionable, because source shapes are complex in general. This is 
visible by residual emission structures after source subtraction. 

We checked the reliability of the listed sources by comparing with the corresponding 
Effelsberg $\lambda$21~cm \citep{rrf+90,rrf+97} and $\lambda$11~cm source lists \citep{frr+90b}. 
We also compared $1\degr \times 1\degr$ $\lambda$6~cm maps centred on each source with 
the corresponding Effelsberg maps at $\lambda$21~cm and $\lambda$11~cm to check their 
reliability further and to identify misidentifications by low-level distortions.  
A number of sources with poor fits shown by large formal errors could be identified as artificial. 
There are well-fitted $\lambda$6~cm sources, which are barely or not at all visible 
in the longer wavelengths surveys.  
 
\subsection{Fit of linearly polarised sources}   

Separating the small-scale emission of polarised sources is not trivial in the presence of significant 
extended polarised emission. We started from the observed Urumqi $\lambda$6~cm Stokes $U$ and $Q$ maps,
which included extended polarised emission structures of up to a few degrees in size. 
We removed the large-scale components by applying the filtering method by \citet{kk+02}, which is 
a modification of the \citet{sr79} filter and separates positive {\it and} negative small-scale 
structures from large-scale emission. The \citet{sr79} filter only separates positive small-scale and
large-scale emission and is only applicable to total intensity or polarised intensity maps.
We filtered the Stokes $U$ and $Q$ survey maps with a filtering beam of $30\arcmin$ and calculated
$PI$-maps via $PI$ = $(U^{2} + Q^{2})^{0.5}$. For all sources identified in total intensity,
we extracted $1\degr \times 1\degr$ large fields centred on the source position in $PI$ and applied
a Gaussian fit, which takes the positive $PI$ noise bias into account. Polarised peak flux 
densities of 10~mJy or higher were accepted. This 
limit is slightly above 3$\times$ the rms-noise of polarised emission. The lowest percentage polarisation 
in our list is 1.8\%. We did not include sources close to the level of instrumental polarisation of the Urumqi 
telescope, which was found to be of about 1\% after cleaning, as discussed by \citet{srh+11}.    
For polarised sources, we also fitted the $U$ and $Q$ maps to calculate the Galactic
polarisation angle $PA$ by $PA$ = 0.5 atan ($U$/$Q$). When either the $U$ or $Q$ value could not be fitted
by a Gaussian, we estimated its amplitude from a scan across the peak of the source.       

\begin{table*}[!htbp]
\caption{{\bf Flux densities of 3832 compact sources at $\lambda$6~cm.}  
The full table is available from the CDS and appended to the electronic version.}
\label{table1}
\centering
\begin{tabular}{rrrr@{\hspace{2mm}}r@{\hspace{2mm}}rr@{\hspace{2mm}}r@{\hspace{2mm}}rrrcl}\hline\hline
\\
  NO. & \multicolumn{1}{c}{L}& \multicolumn{1}{c}{ B} &\multicolumn{3}{c}{RA (2000)} &\multicolumn{3}{c}{DEC(2000)} &  FLUX &  \multicolumn{1}{c}{PEAK FLUX}  &  SIZE  &   ERROR\\
      & \multicolumn{1}{c}{(D)}  &\multicolumn{1}{c}{(D)} &\multicolumn{3}{c}{ (H\hspace{2mm}  M\hspace{2mm}  S)}&\multicolumn{3}{c}{  (D\hspace{2.5mm}  $\arcmin$ \hspace{2.5mm} $\arcsec$)}&  (mJy) &   \multicolumn{1}{r}{(mJy)} &AND P.-ANGLE& CLASS\\\hline
\\
    1&   10.085&   0.742&  18&  5& 10.7& $-$19& 51& 13&      573&     514&        SE     &  222\\ 
    2&   10.161&  $-$0.347&  18&  9& 23.2& $-$20& 19&  3&    63835&   43890&  12.7/10.3/ 28&  2121\\
    3&   10.268&   3.344&  17& 55& 59.5& $-$18& 24& 22&       64&      54&        SE     &  224 \\
    4&   10.311&  $-$0.117&  18&  8& 50.4& $-$20&  4& 29&    15923&   13252&      SE     &  323 \\
    5&   10.324&   0.897&  18&  5&  6.1& $-$19& 34& 10&      124&     124&        PL     &  234 \\
    6&   10.632&  $-$0.390&  18& 10& 31.0& $-$19& 55& 33&    10250&    7654&      SE     &  112 \\
    7&   10.698&  $-$3.168&  18& 21&  5.5& $-$21& 11& 20&      235&     173&  12.2/10.0/ 99&  2121\\
    8&   10.793&  $-$0.875&  18& 12& 39.4& $-$20&  1&  5&      791&     530&  12.2/11.0/ 39&  2241\\
    9&   10.819&   3.481&  17& 56& 39.0& $-$17& 51& 41&       99&      80&        SE     &  246 \\
   10&   10.931&  $-$3.880&  18& 24& 15.6& $-$21& 18& 54&       85&      71&      SE     &  223 \\
   11&   10.971&   3.004&  17& 58& 42.4& $-$17& 58&  3&      404&     404&        PL     &  222 \\
   12&   11.063&  $-$2.597&  18& 19& 39.9& $-$20& 35& 56&      234&     165&  12.7/10.1/ 72&  2231\\
   13&   11.167&  $-$0.356&  18& 11& 29.2& $-$19& 26& 27&     9423&    8036&      SE     &  111 \\
   14&   11.196&   0.118&  18&  9& 47.1& $-$19& 11& 12&     2168&    1635&        SE     &  234 \\
   15&   11.200&  $-$1.087&  18& 14& 16.6& $-$19& 45& 44&     4234&    2107&  14.5/12.5/ 59&  2121\\
   16&   11.235&   3.512&  17& 57& 24.4& $-$17& 29& 10&      150&     112&  11.8/10.3/ 94&  1121\\
   17&   11.303&  $-$4.756&  18& 28& 20.0& $-$21& 23& 25&      360&     322&      SE     &  212 \\
   18&   11.360&   4.330&  17& 54& 42.0& $-$16& 58&  9&      156&     134&        SE     &  222 \\
   19&   11.423&  $-$0.063&  18& 10& 55.2& $-$19&  4& 31&     3586&    2229&  13.7/10.6/116&  1121\\
   20&   11.509&   3.182&  17& 59& 10.6& $-$17& 24& 46&      186&     138&  11.1/10.9/144&  2121\\
   21&   11.555&   0.363&  18&  9& 36.8& $-$18& 45& 14&      605&     540&        SE     &  222 \\
\\
\hline
\end{tabular}
\end{table*}

\section{The source list}

\subsection{Total-intensity data}

We list the parameters of the 3832 catalogued sources in Table~1, which is accessible from the 
CDS in Strasbourg\footnote{http://vizier.u-strasbg.fr} and appended to the electronic version. 

From the Gaussian fit of each source, we calculated its integrated flux density $S_{\rm i}$ assuming 
a Gaussian shape using the fitted peak flux density $S_{\rm p}$ and the major $\theta_{\rm max}$ and the minor 
$\theta_{\rm min}$ axis of the ellipsoid by 
$S_{\rm i} = S_{\rm p} \times (\theta_{\rm max} \times \theta_{\rm min} \times {\rm HPBW}^{-2})$, with HPBW = $9\farcm5$.   
We quote the fitted sizes in Table~1. For Gaussian-shaped sources, the intrinsic size calculates as
source-size = (fitted size$^{2} - {\rm HPBW}^{2})^{0.5}$.
The positional accuracy of the $\lambda$6~cm survey was checked using the positions of NVSS sources as reference
for each survey map \citep{shr+07}. If necessary, they were corrected for a position accuracy of 
better than $1\arcmin$. This error is not included in the positional uncertainty from the Gaussian 
fit listed in column~9 of the source table. The integrated flux density error does not include
the survey scaling error of less than 4\%.
As for the Effelsberg $\lambda$21~cm and $\lambda$11~cm source lists, we use error classes
to quantify the errors from the Gaussian fit.     

\vskip 0.2cm
The source table includes the following data:

\begin{itemize}
\item Column 1: sequential number
\item Columns 2 and 3: Galactic longitude and latitude
\item Columns 4 and 5: Right ascension and declination (J2000)
\item Column 6: Integrated flux density in mJy
\item Column 7: Peak flux density in mJy
\item Column 8: 
                       PL - point-like source: fitted size smaller than $10\arcmin \times 10\arcmin$

                       SE - slightly extended source: fitted size smaller than $11\arcmin \times 11\arcmin$

                       for extended sources:

                       1. number: fitted FWHM along the major axis in arcmin

                       2. number: fitted FWHM along the minor axis in arcmin

                       3. number: Galactic position angle of the source ellipsoid
\item Column 9       : Error class of the fitting procedure:

                       1. digit: positional error in units of $5\arcsec$

                       2. digit: integrated flux density error in units of $5\%$
     
                       3. digit: size error in units of $10\arcsec$
     
                       4. digit: error of the Galactic position angle in units of $1\degr$
\end{itemize}

From the small-diameter SNRs in the $\lambda$6~cm survey, which were discussed by \citet{srr+11},
a number of sources have apparent sizes below the limit of $16\arcmin$ and were thus included in Table~1.
The integrated flux densities from Gaussian fitting and the ring integrations performed
by \citet{srr+11} in general agree within the quoted errors. For a few cases, the
different methods lead to flux density differences, which slightly exceed the quoted errors. The complex surrounding
of G11.1$-$1.0 (source~15), for example, leads to a lower integrated flux density by ring integration (3.40$\pm$0.25~Jy) 
than by the Gaussian fit (4.234$\pm$0.212~Jy). The same is found for G74.9+1.2 (source~1085), where ring integration gives
6.35$\pm$0.35~Jy and the Gaussian fit 7.217$\pm$0.361~Jy. SNR G59.8+1.2 (source~819) has a slightly 
lower Gaussian flux density (1.17$\pm$0.06) than obtained by ring integration (1.43$\pm$0.08~Jy), which, however, 
includes its tail. Some listed sources within the area of SNRs are either compact substructures or unrelated 
background sources. Their flux densities are therefore always below that of the SNR. An example for a Gaussian fit of 
a substructure is G16.8$-$1.1 with 3.91~Jy (source~87) versus 7.39~Jy for the entire object
\citep{srr+11}. Table~1 also contains a few sources in the area of large-diameter SNRs 
studied by \citet{gsh+11a}, where it is not always clear whether they are unrelated background sources or
compact substructures of the SNRs.

\subsection{Polarised sources}

We list 125 polarised $\lambda$6~cm sources in Table~2, which is also available from Vizier at the 
CDS in Strasbourg\footnote{http://vizier.u-strasbg.fr}. Table~2 includes some total-intensity information 
from Table~1 and the following data:

\begin{itemize}
\item Column 1: sequential number from Table~1
\item Columns 2 and 3: Galactic longitude and latitude (L, B) in degrees from Table~1
\item Columns 4 and 5: Integrated ($S_{\rm i}$) and peak flux density ($S_{\rm p}$) in mJy from Table~1
\item Column 6: Polarised peak flux density (PI$_{\rm p}$) in mJy/beam
\item Column 7: Galactic polarisation angle (PA$_{\rm gal}$) in degrees
\item Column 8: Equatorial polarisation angle for epoch 2000 (PA$_{\rm J2000}$) in degrees
\item Column 9: Peak percentage polarisation (PC$_{\rm p}$)
\item Column 10: Remarks: identifications (see text), PC21 (percentage polarisation
at $\lambda$21~cm available)    
\end{itemize}

We list polarised peak flux density (PI$_{\rm p}$), the Galactic and equatorial polarisation angles 
(PA$_{\rm gal}$, PA$_{\rm J2000}$), and the peak percentage polarisation (PC$_{\rm p}$). The polarised flux densities were 
obtained from the $PI$ maps with the same Gaussian fitting software as used for the total-intensity fits. 
The quoted extent from the Gaussian fit of the total-intensity and the much fainter 
polarised emission often differ, so that we quote peak flux densities rather than  
integrated flux density values. To obtain integrated polarised intensity values for extended objects, 
the public survey maps should be used and intensity integrations performed. We have 
excluded the extended sources~84 ({\it l, b = 16.518, $-$3.226} ), 401 ({\it l, b = 36.686, 1.831} ), and 
415 ({\it l, b = 37.429, $-$2.430} ) from Table~2, for which we 
obtained formal percentage polarisations exceeding 35\%. These sources are all located 
towards the inner Galaxy, where polarised emission is rather patchy, so that chance coincidences 
of an unrelated polarised patch with a weak source may happen. These three
sources are all visible in the Effelsberg surveys, but are faint and not listed as discrete sources. 
No further spectral or other information is available for these sources from VizieR at the CDS in Strasbourg. 

In column~9, we have added identifications taken from the CDS, where SNR stands for supernova remnant,
HII for $\ion{H}{II}$ region, PN for planetary nebula, RadGal for radio galaxies, QSO for quasars,
and AGN for active galactic nuclei, where AGN? indicates candidate objects. 
The well-studied 3C sources were also included. For most sources, $\lambda$21~cm percentage 
polarisation is available, which we have marked as PC21 in Table~2. Most of these data come from the 
NVSS \citep{ccg+98}, a few from the CGPS \citep{btj+03} and from the VLA \citep{vbs+11}. 
Polarisation data from \citet{bmv+98} and \citet{ti+80} were also used.          

A few polarised sources listed in Table~2 refer to known SNRs, SNR substructures, or sources in their area. 
The polarised $\lambda$6~cm emission from SNRs in the surveyed region has already been studied by 
\citet{srr+11} and by \citet{gsh+11a}. The polarised emission from G76.9+1.0 could not be determined by 
ring integration \citep{srr+11}, where a Gaussian fit (source~1120) reveals a PC of 4.0\%. For 3C 58 
(source~2052), the Gaussian fit gives 4.7\% instead of 6\% by ring integration. For the SNRs G39.2$-$0.3 (source~439)
and G74.9+1.2 (source~1085), the percentage polarisations obtained from both methods agree.
Sources~2942 ({\it l, b = 179.473, 2.624} ) and 2976 ({\it l, b = 181.440, $-$2.125} ) are located along the shells of 
G179.6+2.0 \citep{gsh+11a} and SNR S147 \citep{xfr+08}, 
respectively. The sources~3185 ({\it l, b = 192.352, 0.372} ) and 3223 ({\it l, b = 194.527, 2.685} ) are located in the direction
of the `Origem Loop', which has recently been shown by \citet{gh+12} to consist of a polarised arc in the north,
most likely a part of an SNR, and $\ion{H}{II}$ regions in the south. 
The extended source~3185 is located near a bright $\ion{H}{II}$-region with detected infrared emission, but 
not identified so far. Source~3223 has a steep spectrum and is most likely extragalactic.   

The sources 3C 154 (source~3070, {\it l, b = 185.592, 4.002} ), 3C 410 (source~1004, {\it l, b = 69.209,$-$3.763} ), 
and also source~3208 ({\it l, b = 193.652, 4.395} ) were observed with the 
Effelsberg 100-m telescope at $\lambda$6~cm \citep{rfr+00}, including polarisation to study
the radio properties of ROSAT X-ray sources. They measured 4\% for 3C 154 
versus 4.3\% (source~3070) in Table~2. For 3C 410 \citet{rfr+00} quote 3\% in agreement with 3.3\% (source 1004).
For source~3208, a steep-spectrum AGN, \citet{rfr+00} measured 9\% at $\lambda$6~cm
versus 5\% in the present list. The Equatorial polarisation angles of the three sources listed in Table~2 agree within
4$\degr$ with those measured by \citet{rfr+00}.

\section{Discussion}

The aim of this paper is to present the $\lambda$6~cm source catalogue in total and polarised 
intensity. The following brief discussion demonstrates the impact of the new catalogue in view of 
existing data sets. Using catalogues at various frequencies from
different telescopes with large differences in angular resolution to calculate source properties, 
such as their spectra, may be a difficult exercise in practice. \cite{vgd+10} present SPECFIND V2.0 
(accessible via CDS/Strasbourg), which is a systematic approach to deriving about 65 10$^{3}$ spectra 
of radio sources. They show examples for the general large scatter in published flux densities, but 
also discuss methods for deriving reliable spectra.

\subsection{Total-intensity data}

\subsubsection{Source statistics}

\begin{figure}
\centering
  \includegraphics[height=8.8cm,angle=270]{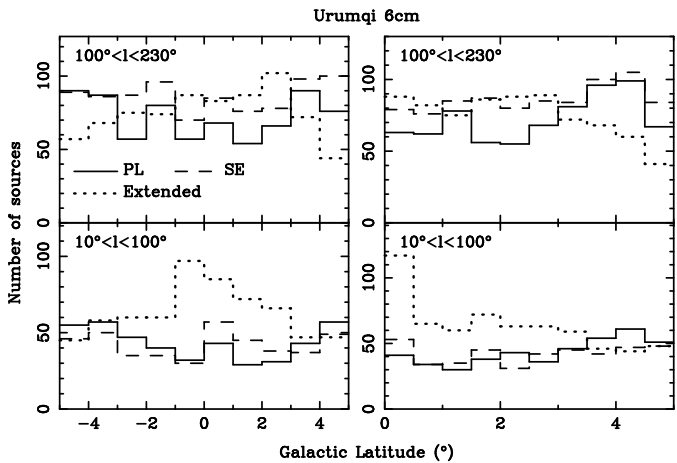}
  \caption{Latitude ({\it left}) and absolute latitude distribution ({\it right}) for the three source 
classes, `PL', `SE', and `Extended', shown for the inner and outer Galaxy.}
\label{fig1}
\end{figure}

The distribution of $\lambda$6~cm sources as a function of Galactic latitude and absolute latitude is shown in Fig.~1.
For Galactic longitudes less than $100\degr$, a concentration of extended sources within $1\degr$ latitude
is visible, which is similar to the Effelsberg $\lambda$11~cm extended source distribution (see \citet{frr+90b}, 
Fig.~4), which is slightly wider in latitude. This is a clear indication 
that extended sources are mostly of Galactic origin. The `PL' and `SE' source distribution is
almost latitude independent, indicating that most of these sources are extragalactic. 
For longitudes above $100\degr$, the distribution shows no clear latitude dependence for all source classes, 
again this agrees with the distribution of $\lambda$11~cm sources (see \citet{frr+90b}, Fig.~5).
We conclude that sources classified as `extended', e.g. with fitted sizes above $11\arcmin$ 
are mostly Galactic. Their density is highest within absolute latitudes of $1\degr$ towards the inner 
Galaxy, where the diffuse Galactic emission also peaks. 

\citet{frr+90b} presented cumulative source counts at $\lambda$11~cm for longitudes from
$100\degr$ to $240\degr$ by counting `PL' and `SE' sources, where the majority are extragalactic. 
The slope of the source count distribution was fitted by $S^{-1.4}$, which is close to $S^{-1.5}$ as 
expected for an isotropic source distribution. The same is found for 
the Effelsberg $\lambda$21~cm `PL' and `SE' sources for the area from $100\degr$ to $230\degr$ as
shown in Fig.~2. The source counts for the Urumqi `PL' and `SE' sources and the GB6 compact $\lambda$6~cm sources,
excluding border sources (B~flag), extended (E~flag) and weak sources with large zero-level (W~flag), and also
sources near a strong source (C~flag) \citep{gsd+96}, are
included in Fig.~2. They show a slightly flatter slope, possibly indicating an increase in the
fraction of Galactic sources at shorter wavelengths or selection effects by confusion. 
The high-latitude compact GB6 source count for latitudes over $10\degr$ is fitted by $S^{-1.5}$  
(Fig.~2). The Galactic plane source numbers drop below the fit for flux densities lower than about 40~mJy for GB6 and 
60~mJy for Urumqi $\lambda$6~cm sources. Below these flux density levels, the catalogues become incomplete. 

\begin{figure}
\centering
  \includegraphics[height=8.5cm,angle=270]{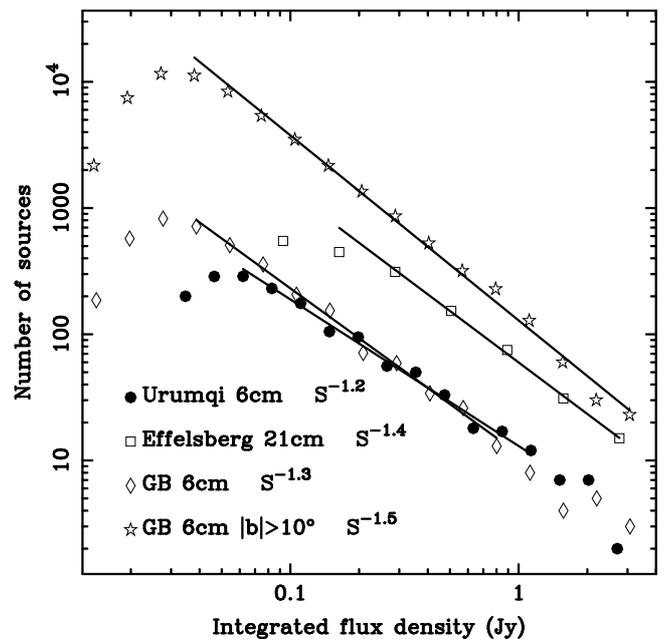}
  \caption{Cumulative source counts in the anti-centre for `PL' and `SE'
sources from the indicated catalogues. We also show the GB6 compact source count 
outside of the Galactic plane.}
\label{fig2}
\end{figure}

\subsubsection{Comparison with the Effelsberg $\lambda$21~cm source list}

The angular resolution of the Sino-German $\lambda$6~cm polarisation survey of $9\farcm5$ matches that of 
the $\lambda$21~cm survey of $9\farcm4$ carried out with the Effelsberg 100-m telescope  \citep{rrf+90,rrf+97}. 
The source lists from both surveys are limited to source sizes of $16\arcmin$.

The minimum peak flux density limit of the $\lambda$21~cm source list for the anti-centre
area for longitudes over $95\fdg5$ \citep{rrf+97} was 79~mJy, while it was 98~mJy 
for longitudes below $95\fdg5$ \citep{rrf+90}.  
For a $\lambda$21~cm peak flux density limit of 98 (79)~mJy, non-thermal sources with 
spectral indices larger than $\alpha$ = $-$1.0 ($-$0.8) ($S\sim \nu{^\alpha}$) should be 
detected and included in the $\lambda$6~cm source list. Fainter sources or sources with a steeper spectrum 
will be missed. The spectral index distribution for Westerbork Northern Sky Survey (WENSS) \citep{rtb+97} sources 
at 327~MHz and NVSS sources
at 1.4~GHz derived from about 186\,000 sources by \citet{zrr+03} shows that about 40\% of 
compact sources have spectra steeper than $\alpha$ = $-$1.0. Thus, a significant fraction of compact $\lambda$21~cm 
sources will be missed at $\lambda$6~cm. On the other hand, flat-spectrum sources with a 
$\lambda$6~cm peak flux density below 90~(70)~mJy will be missed at $\lambda$21~cm.
Some extended sources might be missed, when they slightly exceed the size limit in one catalogue,
but were just below in the other. 

In the latitude limits of $\pm4\degr$ of the Effelsberg $\lambda$21~cm inner Galactic plane survey
and longitudes between $10\degr$ and $95\fdg5$, we find 1127 sources at $\lambda$6~cm, while
the $\lambda$21~cm source list has 827 entries. Among them
673 sources are listed in both catalogues. In the anti-centre area, for longitudes higher than $95\fdg5$, 
we find 1950 sources at $\lambda$6~cm 
compared to 1643 sources at $\lambda$21~cm, where 1414 $\lambda$6~cm sources have a counterpart in the 
$\lambda$21~cm catalogue. 
 
All together, the $\lambda$6~cm source list contains about 20\% more sources than the $\lambda$21~cm source list.
From the mentioned selection effects, we conclude that most of the faint $\lambda$6~cm sources must be flat-spectrum 
synchrotron sources or optically thin thermal sources. This is a significant fraction of sources 
in the Galactic plane at $\lambda$6~cm. 

\begin{figure}[!htbp]
\centering
  \includegraphics[height=8.8cm,angle=270]{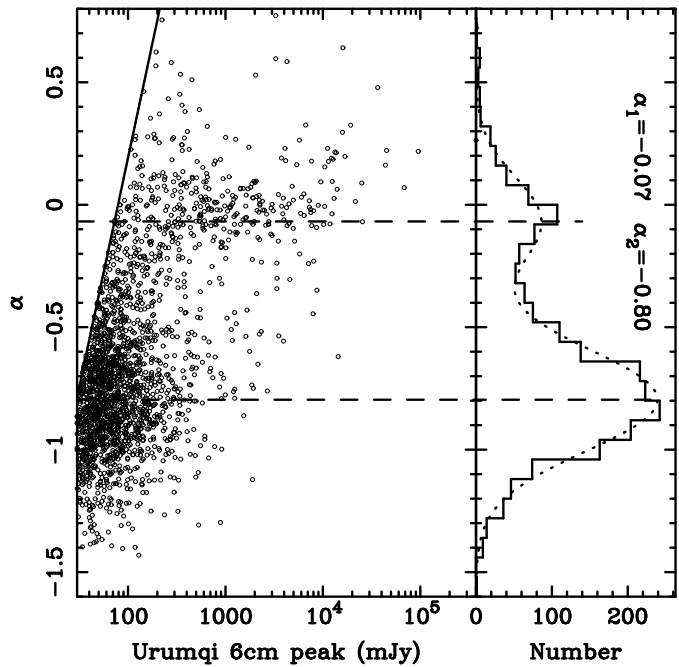}
  \caption{Spectral indices calculated from Effelsberg $\lambda$21~cm and Urumqi $\lambda$6~cm 
peak flux densities. The solid line in the left panel indicates the spectral indices 
derived from the $\lambda$6~cm peak flux densities and the peak flux density limit of 
79~mJy at $\lambda$21~cm. The two dashed lines show the peak locations from double 
Gaussian fitting in dotted lines in the right panel.  
}
\label{fig3}
\end{figure}

The $\lambda$21~cm and $\lambda$6~cm surveys have about the same angular resolution, which allows us to calculate 
spectral indices with  
peak flux densities rather than integrated flux densities, where differences
in the fitted sizes at the two wavelengths decrease the accuracy.
We show the result for sources listed in both catalogues in Fig.~3. The two lines indicate spectral indices 
of $\alpha$ = $-$0.80 and $\alpha$ = $-$0.07, which were derived from a double Gaussian 
fit. Some clustering of sources along these lines is seen. Extragalactic sources have a median spectral index 
of $\alpha\approx-0.9$ \citep{zrr+03}. Fig.~3 shows that most of the strong sources have a flat spectrum 
with $\alpha\approx-0.1$.

\subsubsection{Comparison with the GB6 source list}

The $\lambda$6~cm multi-beam source survey carried out with the former 300-ft Green Bank telescope
\citep{cbs+94} has an angular resolution of $3\farcm6 \times 3\farcm4$.
We compared the GB6 integrated flux densities from the catalogue compiled by \citet{gsd+96} 
with the present source list. The \citet{gsd+96} catalogue includes sources up to $10\farcm5$ in size. 
The present $\lambda$6~cm source list includes sources with sizes up to $13\arcmin$. 
We took into account that for a small number of objects the GB6 catalogue lists two sources because
of its higher angular resolution. Their flux densities were added for the comparison. In cases where the 
two GB6 sources have a positional difference of a few arc-minutes, the Gaussian fit of a single 
extended source has a large error and the residuals are large. We have 
not sorted these few cases out and show the result of the comparison in Fig.~4. 
For many sources, the flux density ratio is close to one, as expected, with an increasing scatter towards lower flux 
densities. There are a number of outliers. Figure~4 shows more sources with significantly higher flux densities in the 
Urumqi catalogue compared to the GB6 list, than for the opposite case. 
For sources with low flux densities, this effect is masked by the general scatter in the
flux density ratio.       

We have marked four `outlier' sources in Fig.~4 that show large differences in flux densities.
We briefly discuss these cases to demonstrate the strengths and limitations of
the different catalogues. All four sources are extended $\ion{H}{II}$ regions \citep{pbd+03} with small-scale
structures measured with interferometers. Source~372 ({\it l, b = 35.075, $-$1.494} ) has an integrated (peak) flux density 
of 5075 (3923)~mJy in the present list and 8692 (1480)~mJy in the GB6 source list. These flux density difference is clearly 
outside the errors of about 10\%. The flux density ratio is 0.58 and thus exceptional low, see Fig.~4. The 
Effelsberg $\lambda$21~cm and $\lambda$11~cm catalogues list 6.27$\pm$0.63~Jy and 6.47$\pm$0.65~Jy 
integrated flux densities, respectively, consistent with the spectrum of an optically-thin 
$\ion{H}{II}$-region. The present $\lambda$6~cm flux density seems to be slightly lower than expected, 
while the GB6 flux density is clearly too high. The much larger GB6 factor to convert peak flux into
integrated flux density resulting from its smaller beam size seems to cause this inconsistency, which is connected to
the uncertainties in the size determination.
The other three sources~421 ({\it l, b = 37.852, $-$0.331} ), 1144 ({\it l, b = 79.284, 0.291} ), and 2429 
({\it l, b = 150.378, $-$1.604} ) 
are `outliers' in the other direction, with a flux density ratio clearly 
exceeding 1 as seen from Fig.~4. The single-dish spectrum of source~421 is inverted up to 8.35~GHz \citep{lmd+00},
which indicates the presence of an optically-thick sub-component within the $\ion{H}{II}$-region.  
The two other $\ion{H}{II}$ regions, sources~1144 and 2429, are optically thin. Numerous small components 
were detected by interferometers, which do not give the correct integrated flux density when summed up. The 
factor to convert peak into integrated flux density is near 2 for both sources and catalogues. 
The maps of these extended sources show a core-halo structure with a compact not always centred core.  
The obtained integrated flux density depends on the beam size to include the entire source. The almost
identical Effelsberg $\lambda$21~cm and Urumqi $\lambda$6~cm beams imply that from these flux densities the 
most reliable spectra are obtained for extended sources.

\begin{figure}
\centering
  \includegraphics[height=9cm,angle=270]{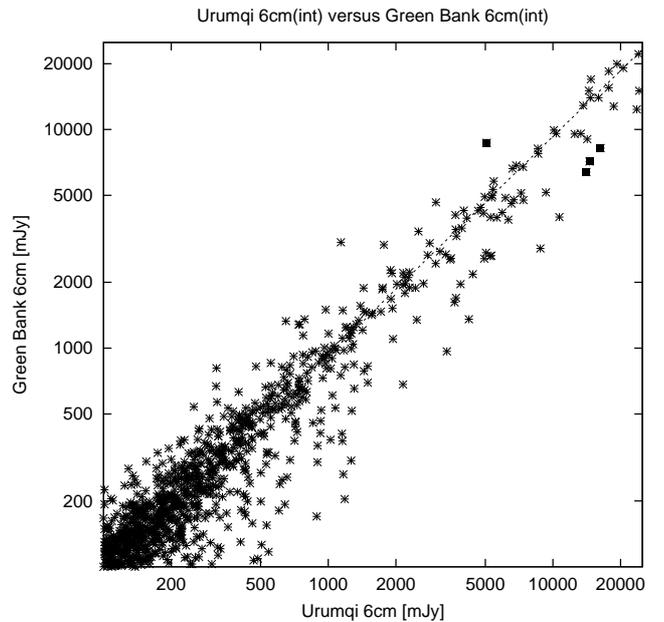}
  \caption{Comparison of integrated flux densities from the present $\lambda$6~cm catalogue with those from
the GB6 source list. Four sources with large flux differences are marked by squares and were discussed in the text.}
\label{fig4}
\end{figure}

\subsection{The polarised sources}

\subsubsection{Distribution along the Galactic plane}

The latitude distribution of polarised $\lambda$6~cm sources in the Galactic plane is nearly uniform.
From the 125 sources, 61 have absolute latitudes below $2\fdg5$ and 64 between $2\fdg5$ and
$5\degr$. The longitude distribution, however, shows an increase in the number density with longitude:
51 sources are between $10\degr$ and $120\degr$ and 74 sources between $120\degr$ and $230\degr$,
which means that depolarisation and confusion is higher towards the inner Galaxy than for its outer region.
For the survey section from $10\degr$ to $60\degr$, \cite{srh+11} showed that the `polarisation horizon' 
is about 4~kpc at $\lambda$6~cm, otherwise the Galaxy is Faraday-thin.       
Among the polarised sources, there are both Galactic and extragalactic sources. Although the 
absolute number of polarised sources is small, 
the uniform distribution and missing concentration of polarised sources towards the Galactic plane 
indicates that the fraction of Galactic sources in the sample is quite small. 
 
We compared the number of polarised $\lambda$21~cm sources from the NVSS with
an angular resolution of 45$\arcsec$ for the area of the $\lambda$6~cm survey observed with a 
$9\farcm5$ beam. Since the majority of polarised sources is extragalactic with a mean non-thermal spectral index around
$\alpha$ = $-$0.9, a polarised $\lambda$21~cm flux density of 30~mJy corresponds to the 
10~mJy polarised source limit at $\lambda$6~cm in case of no depolarisation. 
The number of polarised NVSS sources is
88, where 9 of them are double sources, which were not resolved with the large Urumqi beam. 
Thirty-two of the NVSS sources are between $10\degr$ 
and $120\degr$ and 56 sources between $120\degr$ and $230\degr$ longitude. 
The NVSS numbers are close to those from the Urumqi survey, which indicates that the 
angular resolution and the wavelength are not an important selection effect for sources
with strong polarised emission. The NVSS lists about 20\% more strong polarised sources 
(>30~mJy) for the entire longitude range, but for higher latitudes: 95 sources for +$5\degr$ to +$15\degr$ and 
113 sources for $-15\degr$ to $-5\degr$. 
The NVSS polarised source deficit in the Galactic plane refers entirely to the inner Galaxy, where the 
Galactic plane gets Faraday-thick \citep{srh+11}.       
This indicates that a line-of-sight of several kpc through the Galactic disk is needed 
to cause depolarisation of strong polarised signals on small scales. This changes for fainter polarised NVSS sources,
where a latitude dependence exists. 
Selection effects of confusion with the fluctuating diffuse Galactic emission are more severe and
have some influence on the detection of polarised sources.

\subsubsection{Percentage polarisation}

So far, no systematic survey for polarised sources at $\lambda$6~cm along the Galactic plane
has been available. Thus, we compared the $\lambda$6~cm polarisation data with $\lambda$21~cm data
from the catalogues listed in Sect.~4.2 and indicated in Table~2 as PC21.
The polarised $\lambda$6~cm sources have 88 counterparts at $\lambda$21~cm.
In most cases, the percentage polarisation, PC, increases towards the shorter wavelength
as can be seen from Fig.~5, where $\lambda$6~cm PC is plotted versus $\lambda$21~cm PC.
Faraday rotation depends on $\lambda^{2}$, and thus intrinsically highly polarised sources
are seen to be more depolarised at $\lambda$21~cm than at $\lambda$6~cm. With few exceptions,
the $\lambda$21~cm polarisation data are from high-resolution interferometric data, which 
resolve some single sources in the $\lambda$6~cm catalogue into two polarised components,
but differential Faraday rotation within the sources is not resolved.
The weak effect of Galactic differential Faraday rotation 
towards the inner Galaxy as taken from the source distribution discussed above indicates that the  
increase in the percentage polarisation at $\lambda$6~cm results mainly from decreasing 
internal source depolarisation.

\subsubsection{Identification}

Compared to the number of fitted total-intensity sources, the fraction of 125 Gaussian-fitted
linearly polarised sources is just 3.3\%. One may ask what is special to this small
subgroup of sources that they are polarised, while the majority is not. We have used the VizieR 
service of CDS for source identifications and found 32 extragalactic objects. 
Nine are Galactic and four sources might be SNR substructures or extragalactic sources projected against
SNR shells. From the remaining 80 sources, 68 have spectra steeper than $\alpha$ = $-$0.6, 
3 have spectra flatter than $\alpha$ = $-$0.1. This indicates that more than 80\% of the polarised
sources are extragalactic. Among the nine identified Galactic sources, six are SNRs, one is a planetary 
nebula, and two polarised objects were catalogued as $\ion{H}{II}$ regions.   
In principle, a $\ion{H}{II}$ region may depolarise and/or act as a Faraday screen when
hosting a regular magnetic field component to rotate polarised background emission. This will cause a 
difference to the polarised emission in its surroundings. 
Such objects were discussed by \citet{shr+07}, including modelling, and also in other $\lambda$6~cm survey papers. 

Source~627 ({\it l, b = 50.192, 3.307} ) is identified with the flat-spectrum planetary nebula PK050+31 with an
apparent diameter 
of $2\arcmin$ at 688~pc distance \citep{ssv+08}. The object might be similar to the planetary nebula
Sh 2$-$216 discussed by \citet{ruk+08} and might act as a Faraday screen in the same way as described for
$\ion{H}{II}$ regions above. Because its distance is small, a large fraction of the polarised
Galactic emission is located behind PK050+31 and gets rotated.

\begin{figure}
\centering
  \includegraphics[height=9cm,angle=270]{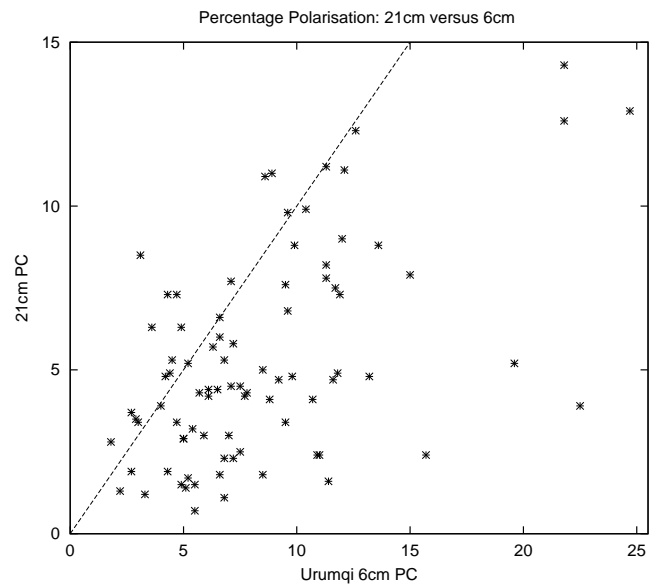}
  \caption{Source percentage polarisation at $\lambda$6~cm versus $\lambda$21~cm.} 
\label{fig5}
\end{figure}

\section{Concluding remarks}

We present a list of 3832 compact sources extracted from the Sino-German $\lambda$6~cm 
polarisation survey of the Galactic plane, where 125 or about 3.3\% of the sample are polarised. 
The $\lambda$6~cm survey complements earlier $\lambda$21~cm and $\lambda$11~cm
surveys from the Effelsberg 100-m telescope with similar angular resolution and sensitivity. 
Most of the listed sources have counterparts at $\lambda$21~cm and $\lambda$11~cm, where the
extension of the wavelength-range up to $\lambda$6~cm allows a more precise spectrum 
determination. It is of interest to determine the spectrum of 
$\ion{H}{II}$ regions at high frequencies, in particular, when they include compact optically-thick 
components. The identification of objects with spinning dust emission, which 
peaks between $\lambda$3~cm and $\lambda$1~cm, critically depends on reliable thermal emission spectra.  
The $\lambda$6~cm source catalogue lists about 20\% more sources than the Effelsberg $\lambda$21~cm
catalogue, which must be faint flat-spectrum sources. Their location in 
the Galactic plane suggests that most of them are faint $\ion{H}{II}$ regions, although confirmation is needed. 

We compared the integrated flux densities from the Urumqi $\lambda$6~cm survey with those
measured with the former Green Bank 300-ft telescope (GB6) with higher angular resolution. 
The integrated flux densities of extended sources seem to be more precise in the present catalogue. 

We found a similar number and distribution of strong polarised NVSS sources at $\lambda$21~cm 
compared to polarised $\lambda$6~cm sources in the Galactic plane, despite their large angular resolution 
difference. The percentage 
polarisation increases from $\lambda$21~cm to $\lambda$6~cm. We conclude that the depolarisation 
properties of compact sources are mainly caused by internal effects, while small-scale Galactic Faraday effects
do not contribute to the depolarisation except for the inner Galaxy with lines-of-sight of several kpc.

\begin{acknowledgements}
We thank Ernst F\"urst for his support of the $\lambda$6~cm
survey project and for critical reading of the manuscript. We acknowledge the help of 
Maja Kierdorf with source fitting and table editing.
We are grateful to the staff of the Urumqi Observatory for qualified 
assistance with the survey observations. We thank  
Otmar Lochner for the construction, installation, and comissioning of the $\lambda$6~cm 
receiver. Maozheng Chen and Jun Ma helped with the receiver installation and maintainance 
during the survey project. The MPG and the NAOC supported the construction of 
the Urumqi $\lambda$6~cm receiving system by special funds. The 
Chinese survey team is supported by the National Natural Science foundation of 
China (10773016, 11303035) and the National Key Basic Research 
Science Foundation of China (2007CB815403). XYG, LX, and XHS acknowledge financial 
support by the MPG, by Richard Wielebinski, and Michael Kramer during their various 
stays at the MPIfR Bonn. XHS was supported by the Australian Research Council 
through grant FL100100114.
This research has made use of the VizieR catalogue access tool, CDS, Strasbourg, France.
\end{acknowledgements}

\begin{longtab}
\bgroup
\footnotesize\ttfamily
 \begin{longtable}{rrrrrrrrrl}
\multicolumn{9}{c}{\bf Table~2: List of compact polarised $\lambda$6~cm sources}\\ 
\\
\hline
\\
No\hspace{1mm} & L\hspace{5mm} & B\hspace{5mm} & S$_{i}$\hspace{2.5mm} & S$_{p}$\hspace{2.5mm} & PI$_{p}\hspace{1.5mm}$ 
& PA$_{gal}$ & PA$_{J2000}$ & PC$_{p}$ & Remark \\
\hspace{1mm} &  (D)\hspace{3.2mm} & (D)\hspace{3.2mm} & (mJy) & (mJy) & (mJy) 
& (D) & (D)\hspace{1.5mm} & (\%) &  \\
\\
\hline
\\ 
   34 & 12.425 &  3.846 &   1281 &  1209 &  76 &  176 &  116 &  6.8 & PC21\\
   35 & 12.447 & -1.123 &   1467 &  1120 &  88 &    8 &  126 &  7.3 & \\
  101 & 18.379 & -3.480 &    172 &   125 &  27 &  128 &   65 & 21.4 & \\
  107 & 18.593 & -2.297 &    189 &   189 &  29 &  142 &   79 & 15.3 & \\
  193 & 24.973 &  4.401 &    486 &   424 &  24 &  109 &   47 &  5.7 & QSO\\
  232 & 27.301 &  3.516 &    134 &   104 &  21 &   37 &  154 & 20.2 & \\
  244 & 28.190 & -3.992 &     76 &    46 &  10 &  165 &  102 & 21.7 & \\
  285 & 30.128 &  1.337 &   2042 &  1866 & 128 &  110 &   47 &  6.8 & RadGal PC21\\
  324 & 32.648 &  4.423 &    224 &   224 &  26 &  135 &   72 & 11.6 & \\
  379 & 35.486 & -4.689 &     64 &    63 &  15 &   97 &   34 & 23.8 & \\
  388 & 36.014 & -2.842 &    436 &   426 &  22 &  129 &   66 &  5.2 & QSO PC21\\
  439 & 39.233 & -0.322 &  10334 &  7100 & 213 &   15 &  132 &  3.0 & SNR G39.2-0.3 \\ 
  476 & 41.599 & -2.924 &    330 &   330 &  21 &    5 &  123 &  6.4 & \\
  537 & 45.405 &  4.149 &    811 &   743 &  41 &    2 &  119 &  5.5 & 3C394 PC21\\
  621 & 49.999 &  1.779 &    113 &    92 &  18 &   27 &  145 & 19.6 & PC21\\      
  627 & 50.192 &  3.307 &    214 &   201 &  13 &   27 &  144 &  6.5 & PN\\
  688 & 53.522 &  3.159 &    434 &   326 &  23 &  144 &   82 &  7.1 & PC21\\
  717 & 54.736 & -0.099 &   1220 &   703 & 152 &  112 &   51 & 21.7 & SNR G54.7-0.1 \\   
  722 & 54.942 &  2.304 &    444 &   385 &  19 &   47 &  165 &  4.9 & PC21\\
  733 & 55.556 &  2.263 &   2300 &  2040 &  37 &  137 &   75 &  1.8 & AGN PC21\\
  742 & 56.093 &  0.108 &    317 &   289 &  17 &    3 &  122 &  5.9 & QSO PC21\\
  775 & 57.829 & -2.354 &    161 &   152 &  15 &  157 &   97 &  9.9 & PC21\\
  796 & 58.766 &  0.650 &    358 &   209 &  12 &    0 &  119 &  5.7 & HII\\
  848 & 61.433 &  4.759 &    223 &   223 &  16 &  107 &   45 &  7.2 & PC21\\
  899 & 63.684 & -2.334 &    472 &   472 &  42 &  148 &   90 &  8.9 & PC21\\
  906 & 64.051 & -4.321 &    235 &   196 &  13 &    0 &  123 &  6.6 & QSO PC21\\
  938 & 65.288 & -4.948 &    224 &   182 &  24 &    9 &  132 & 13.2 & PC21\\
  949 & 65.913 & -4.591 &    252 &   252 &  12 &   84 &   27 &  4.8 & AGN?\\
  980 & 67.748 &  1.831 &    354 &   244 &  56 &  164 &  105 & 23.0 & AGN?\\
  993 & 68.587 & -0.557 &    270 &   230 &  14 &   45 &  168 &  6.1 & PC21\\
  995 & 68.754 &  0.268 &    298 &   187 &  42 &  131 &   73 & 22.5 & PC21\\
  996 & 68.778 &  2.786 &   5282 &  3383 & 537 &   24 &  145 & 15.9 & SNR CTB80 \\
 1004 & 69.209 & -3.763 &   3682 &  3529 & 116 &  175 &  119 &  3.3 & 3C410 PC21\\
 1053 & 72.253 & -0.974 &   1242 &   960 & 131 &  150 &   94 &  13.6 & PC21\\
 1085 & 74.944 &  1.156 &   7217 &  4571 & 221 &  155 &   99 &   4.8 & SNR G74.9+1.2 \\
 1099 & 75.495 & -3.130 &    152 &    78 &  12 &   60 &    7 &  15.4 & \\
 1120 & 76.883 &  0.959 &    714 &   500 &  20 &   14 &  139 &   4.0 & SNR G76.9+0.9 \\
 1124 & 77.225 & -3.237 &    429 &   341 &  22 &  168 &  116 &  6.5 & PC21\\
 1126 & 77.307 &  1.248 &   1075 &   783 &  38 &  178 &  123 &  4.9 & PC21\\
 1227 & 85.711 &  2.038 &    378 &   325 &  14 &  128 &   77 &  4.3 & HII\\
 1299 & 90.300 & -3.782 &    306 &   254 &  18 &  132 &   88 &  7.1 & PC21\\
 1304 & 90.518 &  1.298 &    644 &   496 &  42 &   30 &  163 &  8.5 & 3C468 PC21\\
 1364 & 93.317 &  0.402 &   1499 &  1241 &  54 &   52 &    8 &  4.4 & QSO PC21\\
 1454 & 97.656 & -3.543 &    622 &   552 &  15 &    8 &  151 &  2.7 & QSO PC21\\
 1552 & 103.281 & -3.299 &    784 &   579 &  17 &   68 &   37 &  2.9 & QSO PC21\\
 1637 & 107.449 &  3.694 &    176 &   132 &  13 &   48 &   17 &  9.8 & QSO PC21\\
 1677 & 109.559 &  3.750 &    191 &   157 &  12 &  138 &  111 &  7.6 & \\
 1712 & 112.126 & -4.488 &    273 &   220 &  25 &  160 &  142 & 11.4 & PC21\\
 1771 & 115.736 & -1.082 &    128 &   106 &  12 &   51 &   38 & 11.3 & PC21\\
 1804 & 117.484 &  0.416 &    249 &   174 &  13 &   57 &   46 &  7.5 & PC21\\
 1823 & 118.546 & -1.265 &   2021 &  1771 &  48 &   23 &   15 &  2.7 & QSO PC21\\
 1856 & 120.555 &  1.201 &    774 &   774 &  36 &   25 &   20 &  4.7 & PC21\\
 1867 & 121.212 & -2.325 &    653 &   614 &  19 &   87 &   84 &  3.1 & PC21\\
 1881 & 121.889 & -4.225 &    122 &   104 &  10 &  160 &  158 &  9.6 & PC21\\
 1924 & 124.335 &  4.850 &    275 &   211 &  22 &    0 &    3 & 10.4 & PC21\\
 2052 & 130.719 &  3.089 &  31286 & 25491 & 1210 & 149 &  165 &  4.7 & SNR 3C58 PC21\\
 2078 & 132.073 &  0.202 &    348 &   305 &  29 &  164 &    1 &  9.5 & PC21\\
 2105 & 133.886 &  4.377 &    433 &   433 &  37 &   98 &  122 &  8.5 & AGN? PC21\\
 2116 & 134.509 & -2.006 &    221 &   171 &  12 &  176 &   16 &  7.0 & PC21\\
 2137 & 136.143 &  4.230 &    409 &   285 &  12 &    0 &   27 &  4.2 & QSO\\
 2139 & 136.211 & -0.899 &    984 &   915 &  52 &   47 &   71 &  5.7 & 3C69 PC21\\
 2160 & 137.544 &  3.966 &    399 &   368 &  43 &   46 &   76 & 11.7 & PC21\\
 2211 & 140.120 & -2.840 &    166 &   144 &  17 &  127 &  155 & 11.8 & PC21\\
 2223 & 140.875 &  2.430 &    197 &   168 &  12 &   82 &  115 &  7.1 & PC21\\
 2245 & 142.039 &  0.324 &    123 &    83 &  10 &  113 &  146 & 12.0 & PC21\\
 2276 & 143.496 & -4.378 &    289 &   273 &  18 &   12 &   43 &  6.6 & PC21\\
 2283 & 143.895 & -1.053 &   2830 &  2695 & 137 &   45 &   79 &  5.1 & PC21\\
 2295 & 144.431 &  0.908 &    184 &   109 &  12 &   30 &   66 & 11.0 & QSO PC21\\
 2305 & 144.982 & -4.702 &    121 &   121 &  12 &   28 &   61 &  9.9 & \\
 2311 & 145.087 &  3.019 &    428 &   412 &  41 &   80 &  119 & 10.0 & \\
 2317 & 145.213 & -2.337 &   1002 &   871 &  59 &  122 &  157 &  6.8 & PC21\\
 2336 & 145.953 & -1.681 &    271 &   271 &  15 &    5 &   41 &  5.5 & PC21\\
 2353 & 146.933 & -0.513 &    365 &   264 &  12 &  167 &   25 &  4.5 & QSO PC21\\
 2366 & 147.814 & -3.898 &   1365 &  1245 &  52 &  163 &   19 &  4.2 & PC21\\
 2432 & 150.575 &  4.561 &    598 &   563 &  17 &   82 &  129 &  3.0 & PC21\\
 2440 & 150.857 &  4.536 &    291 &   252 &  27 &   81 &  128 & 10.7 & PC21\\
 2447 & 151.319 &  2.607 &    701 &   688 &  15 &   38 &   83 &  2.2 & QSO PC21\\
 2509 & 154.208 &  1.152 &    271 &   251 &  18 &   28 &   74 &  7.2 & PC21\\
 2519 & 154.677 &  3.377 &    257 &   228 &  15 &   27 &   76 &  6.6 & PC21\\
 2603 & 158.448 & -4.253 &    380 &   301 &  29 &  151 &   16 &  9.6 & PC21\\
 2639 & 160.401 &  0.138 &   2443 &  1929 &  75 &   35 &   85 &  3.9 & 3C129 \\
 2711 & 164.691 &  4.628 &     55 &    55 &  12 &  133 &    9 & 21.8 & PC21\\
 2761 & 167.640 & -1.900 &   2157 &  1892 & 226 &   30 &   83 & 11.9 & 3C134 PC21\\
 2764 & 167.693 &  0.544 &     73 &    46 &  11 &   64 &  118 & 23.9 & \\
 2833 & 172.972 &  2.440 &    274 &   216 &  34 &   40 &   98 & 15.7 & PC21\\
 2857 & 174.369 & -4.453 &    296 &   255 &  29 &   32 &   86 & 11.4 & RadGal\\
 2860 & 174.531 & -1.319 &    526 &   526 &  57 &  168 &   44 & 10.8 & 3C141 \\
 2930 & 178.832 &  2.646 &    410 &   324 &  13 &   36 &   96 &  4.0 & PC21\\
 2942 & 179.473 &  2.624 &    130 &    80 &  16 &   14 &   74 & 20.0 & SNR-Shell \\
 2966 & 180.746 &  0.955 &    389 &   337 &  31 &   39 &   98 &  9.2 & PC21\\
 2976 & 181.440 & -2.125 &    119 &    70 &  17 &  112 &  170 & 24.2 & SNR-Shell\\
 3000 & 182.353 & -0.623 &    349 &   300 &  23 &   72 &  131 &  7.7 & PC21\\
 3011 & 182.766 & -4.942 &    101 &   101 &  14 &   18 &   75 & 13.9 & \\
 3047 & 184.467 &  3.045 &    294 &   271 &  26 &   58 &  119 &  9.6 & \\
 3070 & 185.592 &  4.002 &   1748 &  1541 &  66 &  136 &   18 &  4.3 & 3C154 QSO PC21\\
 3133 & 189.089 & -4.045 &    328 &   302 &  13 &   30 &   89 &  4.3 & PC21\\
 3145 & 189.942 &  1.976 &    115 &    87 &  11 &  111 &  173 & 12.6 & PC21\\
 3148 & 190.126 & -1.654 &    460 &   366 &  10 &   24 &   84 &  2.7 & \\
 3185 & 192.352 &  0.372 &    161 &    71 &  13 &  123 &    4 & 18.3 & Origem Loop\\
 3208 & 193.652 &  4.395 &    447 &   418 &  21 &  164 &   47 &  5.0 & AGN PC21\\
 3223 & 194.527 &  2.685 &    195 &   125 &  12 &  172 &   54 &  9.5 & Origem Loop PC21\\
 3241 & 195.863 & -3.738 &    112 &   112 &  13 &  158 &   39 & 11.6 & PC21\\
 3251 & 196.329 & -2.316 &     99 &    73 &  18 &   56 &  117 & 24.7 & PC21\\
 3254 & 196.587 &  3.197 &   1039 &   914 &  49 &  111 &  174 &  5.4 & QSO\\
 3260 & 197.062 &  0.323 &     84 &    80 &  12 &   26 &   88 & 15.0 & PC21\\
 3262 & 197.146 & -0.860 &    771 &   740 &  65 &  146 &   28 &  8.8 & PC21\\
 3341 & 200.874 &  0.457 &    162 &   124 &  10 &   77 &  139 &  8.1 & \\
 3342 & 200.892 &  3.197 &    127 &   115 &  13 &  158 &   41 & 11.3 & PC21\\
 3364 & 202.184 & -4.314 &    247 &   159 &  18 &   10 &   72 & 11.3 & PC21\\
 3416 & 205.415 & -4.435 &    291 &   291 &  15 &  106 &  168 &  5.2 & PC21\\
 3424 & 205.799 &  4.905 &    182 &   166 &  13 &   48 &  111 &  7.8 & PC21\\
 3477 & 209.632 &  2.761 &     81 &    71 &  11 &  145 &   28 & 15.5 & \\
 3482 & 210.106 & -2.635 &    260 &   213 &  16 &   47 &  110 &  7.5 & PC21\\
 3522 & 212.528 &  3.129 &     61 &    55 &  12 &  153 &   36 & 21.8 & PC21\\
 3533 & 213.130 &  0.553 &    212 &   182 &  22 &   37 &  100 & 12.1 & PC21\\
 3567 & 215.041 &  2.246 &    850 &   787 &  28 &   60 &  123 &  3.6 & QSO PC21\\
 3616 & 217.787 & -3.011 &    301 &   196 &  12 &   66 &  169 &  6.1 & QSO PC21\\
 3622 & 217.958 &  4.249 &    406 &   367 &  20 &   62 &  125 &  5.4 & PC21\\
 3629 & 218.400 &  2.492 &    532 &   460 &  23 &   96 &  159 &  5.0 & QSO PC21\\
 3676 & 221.001 & -4.688 &    187 &   187 &  16 &  176 &   59 &  8.6 & PC21\\
 3766 & 226.062 & -1.767 &    135 &   110 &  11 &   11 &   74 & 10.0 & \\
 3767 & 226.073 &  0.383 &    153 &   119 &  13 &  108 &  170 & 10.9 & PC21\\
 3785 & 227.080 &  1.006 &    750 &   668 &  42 &   47 &  109 &  6.3 & PC21\\  
 3795 & 227.583 & -2.921 &    368 &   332 &  18 &  147 &   30 &  5.4 & QSO\\
 3810 & 228.367 &  2.552 &     78 &    78 &  12 &   59 &  120 & 15.4 & \\ 
\\
\hline\hline
\end{longtable}
\egroup
\end{longtab}

\bibliographystyle{aa}
\bibliography{/homes/xhsun/bibtex}

\begin{thebibliography}{156}
\expandafter\ifx\csname natexlab\endcsname\relax\def\natexlab#1{#1}\fi


\bibitem[{{Broten} {et~al.}(1988){Broten}, {Macleod}, \&  {Vall\'e}}]{bmv+98}
{Broten}, N.~W., {Macleod}, J.~M., \& {Vall\'e}, J.~P. 1988, \apss, 141, 3038

\bibitem[{{Brown} {et~al.}(2003){Brown}, {Taylor}, \&  {Jackel}}]{btj+03}
{Brown}, J.~C., {Taylor}, A.~R., \& {Jackel}, B.~J. 2003, \apjs, 145, 213

\bibitem[{{Condon} {et~al.}(1994){Condon}, {Broderick}, {Seielstad}, {Douglas}, 
\&  {Gregory}}]{cbs+94}
{Condon}, J.~J., {Broderick}, J.~J., {Seielstad}, G.~A., {Douglas}, K., 
\& {Gregory}, P.~C. 1994, \aj, 107, 1829

\bibitem[{{Condon} {et~al.}(1998){Condon}, {Cotton}, {Greisen}, {Yin}, {Perley},
{Taylor}, \&  {Broderick}}]{ccg+98}
{Condon}, J.~J., {Cotton}, W.~D., {Greisen}, E.~W., {et~al.} 1998, \aj, 115, 1693

\bibitem[{{F\"urst} {et~al.}(1990a){F\"urst}, {Reich}, {Reich}, \&
  {Reif}}]{frr+90a}
{F\"urst}, E., {Reich}, W., {Reich}, P., \& {Reif}, K. 1990a, \aaps, 85, 691

\bibitem[{{F\"urst} {et~al.}(1990b){F\"urst}, {Reich}, {Reich}, \& {Reif}}]{frr+90b}
{F\"urst}, E., {Reich}, W., {Reich}, P., \& {Reif}, K. 1990b, \aaps, 85, 805

\bibitem[{{Gao} {et~al.}(2011a){Gao}, {Han}, {Reich}, {Reich}, {Sun}, \&
  {Xiao}}]{ghr+11b}
{Gao}, X.~Y., {Han}, J.~L., {Reich}, W., {et~al.} 2011a, \aap, 529, A159

\bibitem[{{Gao} {et~al.}(2011b){Gao}, {Sun}, {Han}, {Reich},  
  {Reich} \& {Wielebinski}}]{gsh+11a}
{Gao}, X.~Y., {Sun}, X.~H., {Han}, J.~L., {et~al.} 2011b, \aap, 532, A144

\bibitem[{{Gao} {et~al.}(2010){Gao}, {Reich}, {Han}, {Sun}, {Wielebinski},
  {Shi}, {Xiao}, {Reich}, {F{\"u}rst}, {Chen}, \& {Ma}}]{grh+10}
{Gao}, X.~Y., {Reich}, W., {Han}, J.~L., {et~al.} 2010, \aap, 515, A64

\bibitem[{{Gao} \& {Han}(2013){Gao} \& {Han}}]{gh+12}
{Gao}, X.~Y., \& {Han}, J.~L. 2013, \aap, 551, A16

\bibitem[{{Gregory} {et~al.}(1996){Gregory}, {Scott}, {Douglas}, 
\& {Condon}}]{gsd+96}
{Gregory}, P.~C., {Scott}, W.~K., {Douglas}, K., \& {Condon}, J.~J. 1996, \apjs, 103, 427

\bibitem[{{Han} {et~al.}(2013){Reich}, {Sun}, {Gao}, {Xiao}, {Shi}, {Reich},  
  {Reich} \& {Wielebinski}}]{hrs+13}
{Han}, J.~L., {Reich}, W., {Sun}, X.~H., {et~al.} 2013, Int. Journal of Modern Physics:
Conference Series, in press (arXiv 1202.1875)

\bibitem[{{Haslam}(1974)}]{has74}
{Haslam}, C.~G.~T. 1974, \aaps, 15, 333

\bibitem[{{Kothes} \& {Kerton}(2002)}]{kk+02}
{Kothes}, R., \& {Kerton}, C.~R. 2002, \aap, 390, 337

\bibitem[{{Langston} {et~al.}(2000){Langston}, {Minter}, {D'Addario}, {Eberhardt}, {Koski}, 
 \&  {Zubor}}]{lmd+00}
{Langston}, G., {Minter}, A., {D'Addario}, L., {et~al.} 2000, \apj, 119, 2801

\bibitem[{{Paladini} {et~al.}(2003){Paladini}, {Burigana}, {Davies}, {Maino}, {Bersanelli}, 
{Cappellini}, {Platania} \&  {Smoot}}]{pbd+03}
{Paladini}, R., {Burigana}, C., {Davies}, R.~D., {et~al.} 2003, \aap, 397, 213

\bibitem[{{Ransom} {et~al.}(2008){Ransom}, {Uyan{\i}ker}, {Kothes}, \&
  {Landecker}}]{ruk+08}
{Ransom}, R.~R., {Uyan{\i}ker}, B., {Kothes}, R., \& {Landecker}, T.~L. 2008,
  \apj, 684, 1009

\bibitem[{{Reich} {et~al.}(1997){Reich}, {Reich}, \& {F\"urst}}]{rrf+97}
{Reich}, P., {Reich}, W., \& {F\"urst}, E. 1997, \aaps, 126, 413

\bibitem[{{Reich} {et~al.}(1984){Reich}, {F\"urst}, {Steffen}, {Reif}, 
\& {Haslam}}]{rfs+84}
{Reich}, W., {F\"urst}, E., {Steffen}, P., {Reif}, K., \& {Haslam}, C.~G.~T. 
1984, \aaps, 58, 197

\bibitem[{{Reich} {et~al.}(1990a){Reich}, {F\"urst}, {Reich}, \&
  {Reif}}]{rfr+90}
{Reich}, W., {F\"urst}, E., {Reich}, P., \& {Reif}, K. 1990a,
  \aaps, 85, 633

\bibitem[{{Reich} {et~al.}(1990b){Reich}, {Reich}, \&
  {F\"urst}}]{rrf+90}
{Reich}, W., {Reich}, P., \& {F\"urst}, E. 1990b, \aaps, 83, 539

\bibitem[{{Reich} {et~al.}(2000){Reich}, {F\"urst}, {Reich}, {Kothes}, 
{Brinkmann}, \& {Siebert}}]{rfr+00}
{Reich}, W., {F\"urst}, E., {Reich}, P., {et~al.}  2000, \aap, 363, 141

\bibitem[{{Rengelink} {et~al.}(1997){Rengelink}, {Tang}, {de~Bruyn}, {Miley}, 
{Bremer}, {R\"ottgering} \& {Bremer}}]{rtb+97}
{Rengelink}, W., {Tang}, Y., {de~Bruyn}, A.~G., {et~al.}  1997, \aaps, 124, 259

\bibitem[{{Sofue} \& {Reich}(1979){Sofue}, \& {Reich}}]{sr79}
{Sofue}, Y., \& {Reich}, W. 1979, \aaps, 38, 251

\bibitem[{{Stanghellini} {et~al.}(2008){Stanghellini}, {Shaw}, \& 
{Villaver}}]{ssv+08}
{Stanghellini}, L., {Shaw}, R.~A., \& {Villaver}, E. 2008, \apj, 689, 194

\bibitem[{{Sun} {et~al.}(2007){Sun}, {Han}, {Reich}, {Reich}, {Shi},
  {Wielebinski}, \& {F{\"u}rst}}]{shr+07}
{Sun}, X.~H., {Han}, J.~L., {Reich}, W., {et~al.} 2007, \aap, 463, 993

\bibitem[{{Sun} {et~al.}(2011a){Sun}, {Reich}, {Han}, {Reich}, {Wielebinski},
  {Wang}, \& {M{\"u}ller}}]{srh+11}
{Sun}, X.~H., {Reich}, W., {Han}, J.~L., {et~al.} 2011a, \aap, 527, A74

\bibitem[{{Sun} {et~al.}(2011b){Sun}, {Reich}, {Reich}, {Xiao}, {Gao} 
\& {Han},}]{srr+11}
{Sun}, X.~H., {Reich}, P., {Reich}, W., {et~al.} 2011b, \aap, 536, A83

\bibitem[{{Tabara} \& {Inoue}(1980)}]{ti+80}
{Tabara}, H., \& {Inoue}, M. 1980, \aaps, 39, 379

\bibitem[{{Van Eck} {et~al.}(2011){Van Eck}, {Brown}, {Stil}, {Rae}, {Mao},
  {Gaensler}, {Shukurov}, {Taylor}, {Haverkorn}, {Kronberg}, \&
  {McClure-Griffiths}}]{vbs+11}
{Van Eck}, C.~L., {Brown}, J.~C., {Stil}, J.~M., {et~al.} 2011, \apj, 728, 97

\bibitem[{{Vollmer} {et~al.}(2010){Vollmer}, {Gassmann}, {Derri\'ere}, {Boch}, {Louys},
  {Bonnarel}, {Dubois}, {Genova}, \& {Ochsenbein}}]{vgd+10}
{Vollmer}, B., {Gassmann}, B., {Derri\'ere}, S., {et~al.} 2010, \aap, 511, A53

\bibitem[{{Xiao} {et~al.}(2008){Xiao}, {F{\"u}rst}, {Reich}, \& {Han}}]{xfr+08}
{Xiao}, L., {F{\"u}rst}, E., {Reich}, W., \& {Han}, J.~L. 2008, \aap, 482, 783

\bibitem[{{Xiao} {et~al.}(2011){Xiao}, {Han}, {Reich}, {Sun}, {Wielebinski},
  {Reich}, {Shi}, \& {Lochner}}]{xhr+11}
{Xiao}, L., {Han}, J.~L., {Reich}, W., {et~al.} 2011, \aap, 529, A15

\bibitem[{{Zhang} {et~al.}(2003){Zhang}, {Reich}, {Reich}, \& {Wielebinski}}]{zrr+03}
{Zhang}, X., {Reich}, W., {Reich}, P., \& {Wielebinski}, R. 2003, \aap, 404, 57

\end{thebibliography}
\end{document}